\documentclass[11pt,preprintnumbers,aps,amssymb,nofootinbib,amsmath,superscriptaddress]
{revtex4}
\usepackage{epsfig,epsf}
\usepackage{bm} 
\newcommand{\beq}{\begin{equation}}
\newcommand{\beql}[1]{\begin{equation}\label{#1}}
\newcommand{\eeq}{\end{equation}}
\def\bal#1\eal{\begin{align}#1\end{align}}
\newcommand{\eq}[1]{(\ref{#1})}
\newcommand{\fig}[1]{Fig.~\ref{#1}}

\newcounter{topiccounter}
\renewcommand{\b}[1]{{\bm #1}} 
\newcommand{\unit}[1]{\hat {{\bm #1}}} 

\begin{document}

\title{Time and space dependence of electromagnetic field in relativistic heavy-ion collisions }

\author{Kirill Tuchin}

\affiliation{
Department of Physics and Astronomy, Iowa State University, Ames, IA 50011}

\date{\today}


\begin{abstract}

Exact analytical solution for the space-time evolution of electromagnetic field in electrically conducting nuclear matter produced in heavy-ion collisions is discussed. 
It is argued that the parameter that controls the strength of the matter effect on the field evolution is $\sigma\gamma b$, where  $\sigma$ is electrical conductivity, $\gamma$ is the Lorentz boost-factor and $b$ is the characteristic transverse size of the matter. When this parameter is  of the order one or larger, which is the case at RHIC and LHC,  space-time dependence of electromagnetic field is completely different form that in vacuum.  

\end{abstract}

\maketitle


In relativistic heavy-ion collisions, production of valence quarks in the central rapidity region (baryon stopping) is suppressed \cite{Itakura:2003jp}. Hence $Z$ valence quarks of each nucleus continue to travel after heavy-ion collision along the straight lines in opposite directions. These valence quarks carry total electric charge $2Ze$ that creates electromagnetic field in the interaction region. 
Unlike the valence quarks, gluons and sea quarks are produced mostly in the central rapidity region, i.e.\ in a plane perpendicular to the collision axis. It has been argued in \cite{Landau:1953gs,Belenkij:1956cd} that high multiplicity events in heavy-ion collisions can be effectively described using relativistic hydrodynamics. In particular, matter produced in heavy-ion collisions can be characterized by a few transport coefficients. This approach has enjoyed a remarkable phenomenological success (see e.g.\ \cite{Kolb:2003dz}). Since sea quarks carry electric charge, electromagnetic field created by valence quarks depends on the permittivity $\epsilon$, permeability $\mu$ and conductivity $\sigma$ of the produced matter.  

Consider electromagnetic field created by a point charge $e$ moving along the positive $z$-axis with velocity $v$. It is governed by Maxwell equations:
\begin{align}
&\b\nabla \cdot \b B=0\,, &\b \nabla\times \b E= -\frac{\partial\b B}{\partial t}\,,\label{a4}\\
& \b \nabla \cdot \b D= e\delta(z-vt)\delta(\b b)\,, & \b\nabla \times\b H = \frac{\partial \b D}{\partial t}+ \sigma \b E + ev\unit z \delta(z-vt)\delta(\b b)\,,\label{a5}
\end{align}
where $\b r= z\unit z + \b b$ (such that $\b b\cdot \unit z=0$) is the position of the observation point. Performing Fourier transform 
\beql{a7}
\b E(t,\b r)= \int_{-\infty}^\infty \frac{d\omega}{2\pi} \int_{-\infty}^\infty \frac{dk_z}{2\pi}\int \frac{d^2k_\bot}{(2\pi)^2}e^{-i\omega t + ik_z z+ i\b k_\bot\cdot \b b}\,\b E_{ \omega \b k}\,,\quad \text{etc}\,,
\eeq
we get 
\begin{align}
&\b k \cdot \b B_{ \omega \b k}=0\,, &\b k\times \b E_{ \omega \b k}= \mu\omega\b H_{ \omega \b k}\,,\label{a9}\\
& \epsilon\b k \cdot \b E_{ \omega \b k}= -2i\pi e\delta(\omega-k_zv)\,, & \b k\times\b H_{ \omega \b k} = -\omega\tilde\epsilon\b E_{ \omega \b k}  -2\pi i ev\unit z \delta(\omega-k_zv)\,,\label{a11}
\end{align}
where  $\tilde \epsilon = \epsilon+ i\sigma/\omega$. Solution to these equations reads (see e.g.\ \cite{Jackson})
\begin{align}
& \b H_{ \omega \b k}= -2\pi i e v\frac{\b k\times \unit z}{\omega^2\tilde\epsilon\mu-\b k^2}\delta(\omega-k_zv)\,, 
& \b E_{ \omega \b k}= -2\pi i e\frac{\omega \mu v \unit z - \b k/\epsilon}{\omega^2\tilde\epsilon\mu-\b k^2}\delta(\omega-k_zv)\,.\label{a13}
\end{align}
Substituting \eq{a13} into \eq{a7} it is possible to take integral over $\b k$. However, integration over $\omega$ cannot be done in general form, because the dispersion  relations $\epsilon(\omega)$, $\mu(\omega)$ depend on the matter properties.

Later time dependence of electromagnetic field is determined by a singularity of \eq{a13} in the plane of complex $\omega$ that has smallest imaginary part. We assume that the leading singularity is determined by electrical conductivity. (This gives a conservative estimate of the matter effect). Therefore, we adopt a simple model $\epsilon=\mu=1$, i.e.\ neglect polarization and magnetization response of nuclear matter, but take into account its finite electrical conductivity.  Plugging \eq{a13} into \eq{a7} we take first  trivial $k_z$-integral. Integration over $\omega$ for positive values of $x_-=t-z/v$ is done by closing the integration contour over the pole in the lower half-plane of complex $\omega$. In the relativistic limit $\gamma = 1/\sqrt{1-v^2}\gg 1$  the result  is \cite{Tuchin:2013ie} \begin{align}
& 
\b H(t, \b r) = H(t,\b r)\unit \phi= \frac{e}{2\pi\sigma}\unit \phi\int_0^\infty \frac{J_1(k_\bot b)k_\bot^2}{\sqrt{1+\frac{4k_\bot ^2}{\gamma^2\sigma^2}} }\exp\left\{ \frac{1}{2}\sigma\gamma^2x_-\left( 1- \sqrt{1+\frac{4k_\bot^2}{\gamma^2\sigma^2}} \right)\right\}\,dk_\bot\,,\label{a15}\\
&E_z(t, \b r)= \frac{e}{4\pi}\int k_\bot J_0(k_\bot b)\frac{1-\sqrt{1+\frac{4k_\bot^2}{\gamma^2\sigma^2}} }{\sqrt{1+\frac{4k_\bot^2}{\gamma^2\sigma^2}} }
\exp\left\{ \frac{1}{2}\sigma\gamma^2x_-\left( 1- \sqrt{1+\frac{4k_\bot^2}{\gamma^2\sigma^2}} \right)\right\}\,dk_\bot\,, \label{a16}\\
& \b E_\bot(t,\b r)= H(t, \b r)\unit r\,,\label{a17}
\end{align}
where $\unit r$ and $\unit \phi$ are unit vectors of polar coordinates in transverse plane $x,y$. Electromagnetic field is a function of $\b r-\b r'$, where $\b r$ and $\b r'= vt\unit z$ are the positions of the observation point and the moving charge correspondingly. 
In fact, it depends only on distances $z-vt= -vx_-$ and $b$. 

\begin{figure}[ht]
\begin{tabular}{cc}
      \includegraphics[height=4.5cm]{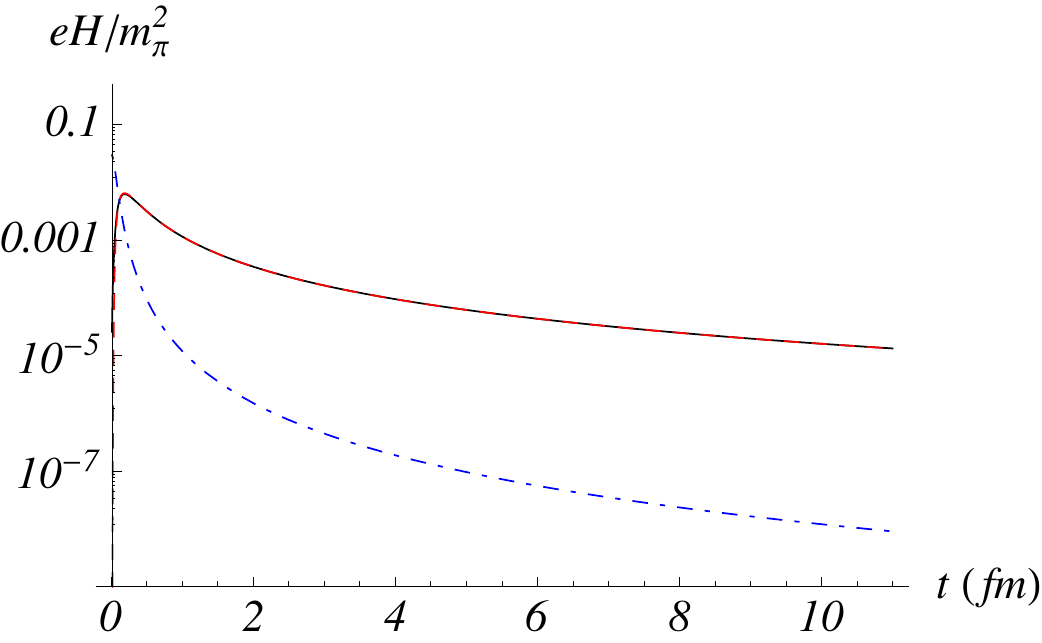} &
      \includegraphics[height=4.5cm]{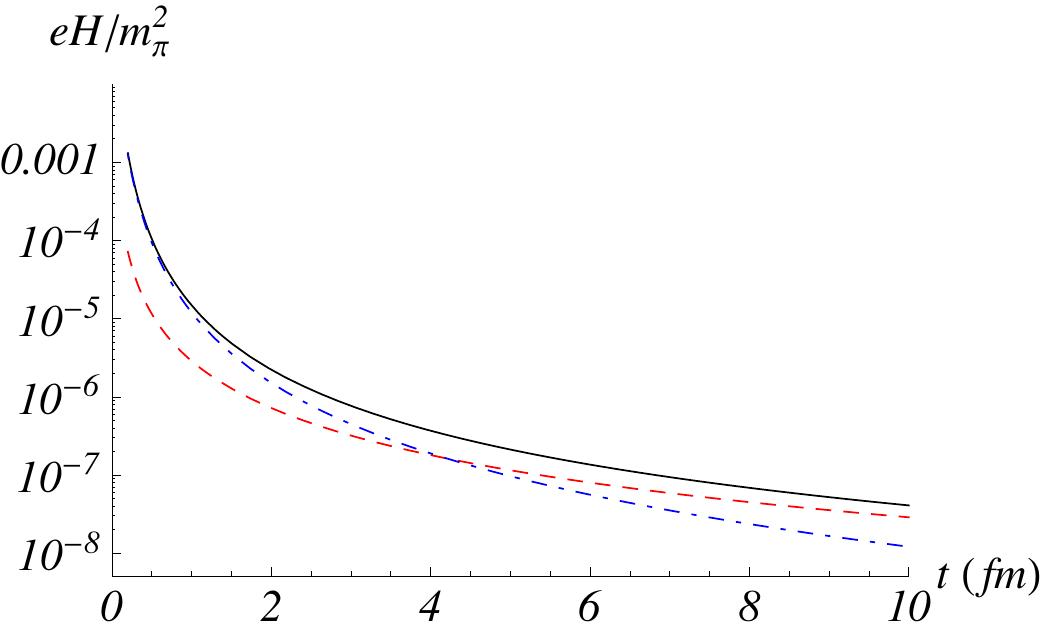}\\
      $(a)$ & $(b)$ 
      \end{tabular}
  \caption{(Color online). Time evolution of magnetic field created by a point unit charge at $z=0$, $b=7.4$~fm, $\gamma =100$ and (a) $\sigma = 5.8$~MeV, (b) $\sigma= 0.01$~MeV. Black solid line is numerical computation of \eq{a15}, red dashed line is  ``diffusion" approximation \eq{a21}, blue dash-dotted line is a solution in free space.  }
\label{fig:2}
\end{figure}

Eqs.~\eq{a15}--\eq{a17} have two instructive limits depending on the value of parameter $\gamma\sigma b$ that appears in the exponents once we notice that $k_\bot\sim 1/b$.  If $\gamma\sigma b\ll 1$, then, after a simple integration, \eq{a15}--\eq{a17} reduce to the boosted Coulomb potential in free space:
\begin{align}\label{a19}
&\b E= \frac{e\gamma}{4\pi}\frac{\b b-v x_- \unit z}{(b^2+\gamma^2v^2 x_-^2)^{3/2}}\,,
&\b H= \frac{e\gamma}{4\pi}\frac{v\unit \phi}{(b^2+\gamma^2v^2 x_-^2)^{3/2}}
\end{align}
This is the solution discussed in \cite{Kharzeev:2007jp}. In the opposite limit $\gamma\sigma b\gg 1$, we expend the square root in \eq{a15},\eq{a16} and derive
\begin{align}\label{a21}
&E_r=H_\phi = \frac{e}{2\pi}\frac{b\sigma}{4x_-^2}e^{-\frac{b^2\sigma}{4x_-}}\,,
& E_z=-\frac{e}{4\pi}\frac{x_--b^2\sigma/4}{\gamma^2 x_-^3}e^{-\frac{b^2\sigma}{4x_-}}\,.
\end{align}
This is the solution pointed out in \cite{Tuchin:2010vs}. Notice that the electromagnetic field in  \eq{a19} drops as $1/x_-^3$ at late times, whereas in conducting matter only as $1/x_-^2$. At RHIC $\gamma=100$, $\sigma\approx 5.8$~MeV  \cite{Ding:2010ga,Aarts:2007wj}. For $b=7$~fm we estimate $\gamma\sigma b= 19$, hence the field is given by the ``diffusive" solution \eq{a21}.  This argument is augmented by numerical calculation presented in \fig{fig:2}. In \fig{fig:2}(a) we plot the result of numerical integration in \eq{a15} for $\sigma\approx 5.8$~MeV and compare it with the asymptotic solutions \eq{a19} and \eq{a21}. It is seen that \eq{a21} completely  overlaps with the exact solution at all times, except at $t<0.1$~fm (not seen in the figure). To illustrate what happens at  $\gamma\sigma b\ll 1$, we plotted in \fig{fig:2}(b) the same formulas as in \fig{fig:2}(a) calculated at artificially reduced conductivity $\sigma\approx 0.01$~MeV. One can clearly observe that at early time matter plays little role in the field time-evolution which follows \eq{a19}, whereas at later time Foucault currents eventually slow down magnetic field decline, which then follows \eq{a21}. This conclusion supports our previous results \cite{Tuchin:2010vs,Tuchin:2013ie} and disagree with the recent claims made in \cite{McLerran:2013hla}.

Electromagnetic field of a charge moving at distance $\b b'$ in the positive $z$-direction with velocity $v$  is given by \eq{a15}--\eq{a17} with $\b b$ replaced by $\b b-\b b'$. Denote it $\b H(x_-,|\b b-\b b'|)$, etc. In the laboratory frame, all charges in a nucleus  have approximately  same longitudinal coordinate $z'=vt$, hence the same $x_-= -(z-z')/v$. 
Therefore, electromagnetic field of  relativistic nucleus can be calculated as 
\beql{a25}
\b H_Z(x_-, \b b)= \int \rho (\b r') \b H(x_-,|\b b-\b b'|) d^3r'= \int  2\sqrt{R_A^2-b'^2}\rho\,\b H(x_-,|\b b-\b b'|)d^2b'\,, \quad\text{etc.},
\eeq
where $\rho= Z/(\frac{4}{3}\pi R_A^3)$ is the nuclear density, $R_A$ is the nuclear radius and we used the fact the $\rho dz'$ is boost-invariant (in $z$-direction).\footnote{We neglect fluctuations of nucleon positions that can also give important contributions to electromagnetic field \cite{Bzdak:2011yy}.   } 
Electromagnetic field of a nucleus moving in the negative $z$-direction is given by \eq{a25} with $x_-$ replaced by  $x_+= t+z/v$. 

\begin{figure}[ht]
      \includegraphics[height=4.5cm]{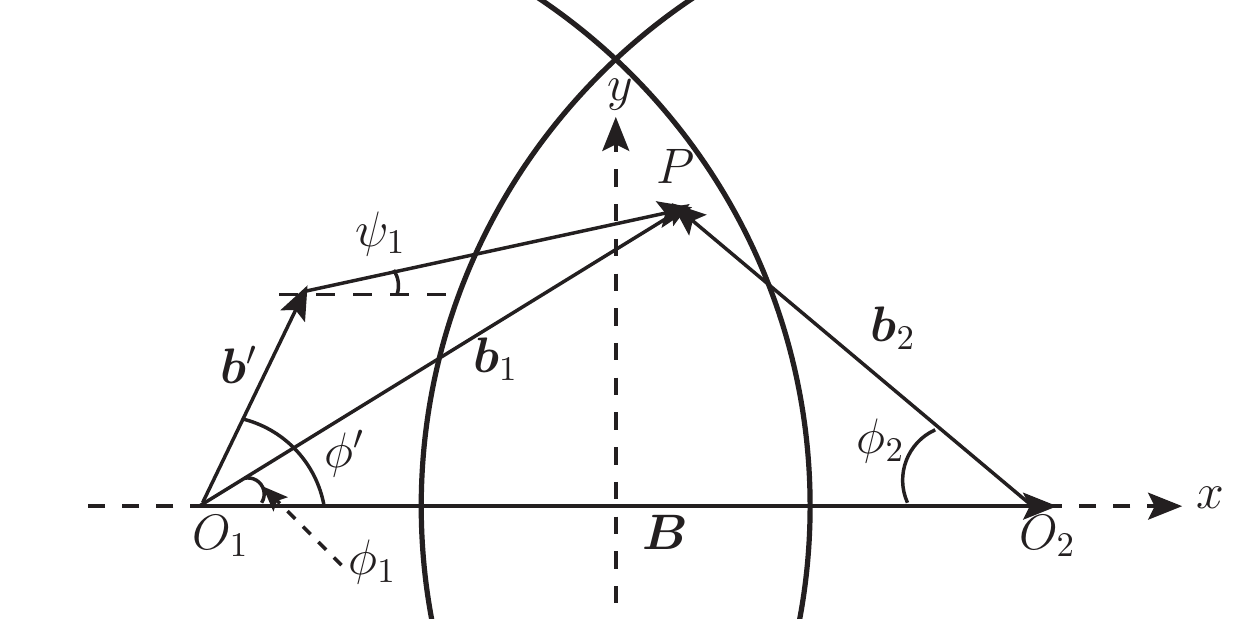}   
      \caption{Transverse plane geometry of heavy-ion collision. Thick lines depict nuclear boundaries. $O_1$ and $O_2$ are nuclear centers. $\b B$ is impact parameter, $\b b_1$ and $\b b_2$ are positions of the observation point $P(x,y)$ with respect to the nuclear centers. $\b b'$ is a position of an elementary charge in nucleus 1.    }
\label{fig:1}
\end{figure}
Consider now total electromagnetic field of two nuclei. Geometry of a heavy-ion collision in transverse plane is depicted in \fig{fig:1}.  Magnetic field at point $P$ with coordinates $x,y$ is directed along the azimuthal angle direction  $\unit\phi= -\sin\psi_1\unit x+\cos\psi_1\unit y$, where $\psi_1$ is the angle between the vector $\b b_1-\b b'$ and $x$-axis, which can be related to vectors $\b b_1$ and $\b b_2$ as follows:
\beql{a27}
\cos\psi_1= \frac{(\b b_1-\b b')\cdot \unit x}{|\b b_1-\b b'|}= \frac{b_1\cos\phi_1-b'\cos\phi'}{\sqrt{b_1^2+b'^2-2b_1b'\cos(\phi'-\phi_1)}}\,.
\eeq
 The transverse component of electric field has radial direction $\unit r= \cos\psi_1\unit x+\sin\psi_1\unit y$. The final expression for the field of a nucleus is 
 \begin{align}
 \b H_Z(x_-, \b b_1)=& \int  2\sqrt{R_A^2-b'^2}\rho\, H(x_-,|\b b_1-\b b'|)(-\sin\psi_1\unit x+\cos\psi_1\unit y)d^2b'\,,\label{a31}\\
 \b E_{Z}(x_-, \b b_1)= &\int  2\sqrt{R_A^2-b'^2}\rho\, \left[ H(x_-,|\b b_1-\b b'|)(\cos\psi_1\unit x+\sin\psi_1\unit y)\right.\nonumber\\
 &\left. + E_z(x_-,|\b b_1-\b b'|) \unit z \right] d^2b' \,,\label{a32}
 \end{align}
 with $\psi_1$ given by \eq{a27} and $H$, $E_z$ by \eq{a15}--\eq{a17}. Similar expressions hold for the other nucleus. The total electromagnetic field of two nuclei is given by
 \bal\label{a35}
\bm{ \mathcal H}(t,z,\b b_1,\b b_2)= \b H_Z(x_-,\b b_1)+ \b H_Z(x_+,\b b_2)\,,\quad \bm{\mathcal  E}(t,z,\b b_1,\b b_2)= \b E_Z(x_-,\b b_1)+ \b E_Z(x_+,\b b_2)\,.
\eal

\begin{figure}[ht]
      \includegraphics[height=4.5cm]{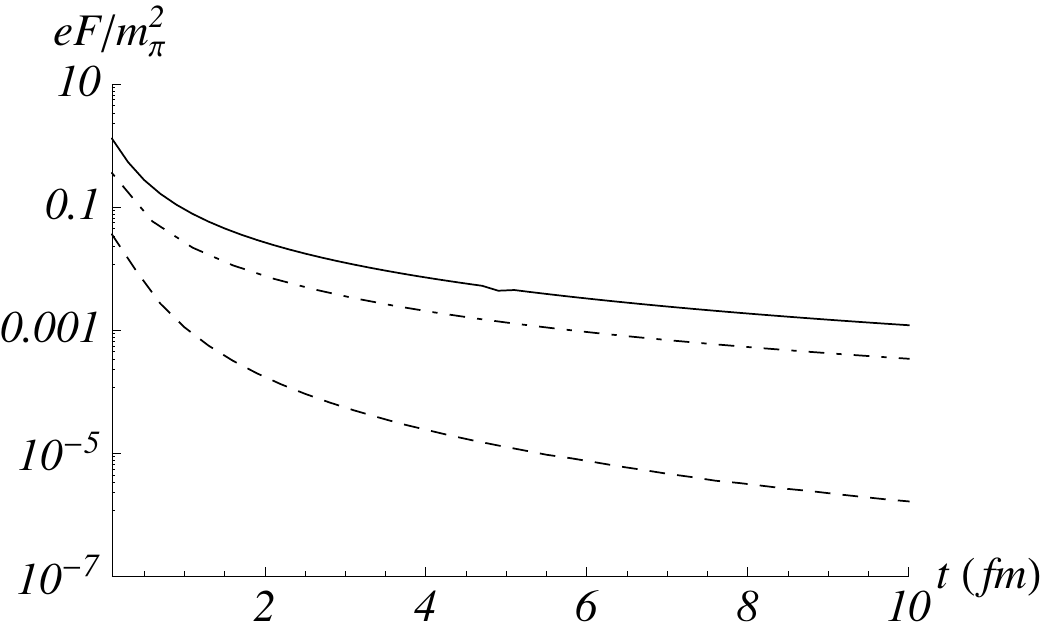} 
  \caption{Time dependence of total electromagnetic magnetic field $F$ at mid-rapidity $z=0$, $\gamma=100$, $B=7$~fm, $t=2$~fm. Solid line: $F=H_y$ at $x=y=0$, dashed line $F=-H_x$ at $x=y=1$~fm, dashed-dotted line $F=-E_y$ at $x=y=1$~fm. }
\label{fig:3}
\end{figure}

In practice one would like to know the electromagnetic field at a given impact parameter 
$\b B=\b b_1-\b b_2 $ as a function of time $t$ and coordinates $x,y,z$ defined in a symmetric way shown in  \fig{fig:1}. This is accomplished using the following equations
\bal\label{a37}
\tan\phi_{1,2}= \frac{y}{x\pm B/2}\,,\quad b_{1,2}= \sqrt{(x\pm B/2)^2+y^2}\,.
\eal
Time dependence of total magnetic field is shown in \fig{fig:3}.   As expected late time dependence of all components is the same and governed by \eq{a21}.   

\begin{figure}[ht]
\begin{tabular}{cc}
      \includegraphics[height=5.5cm]{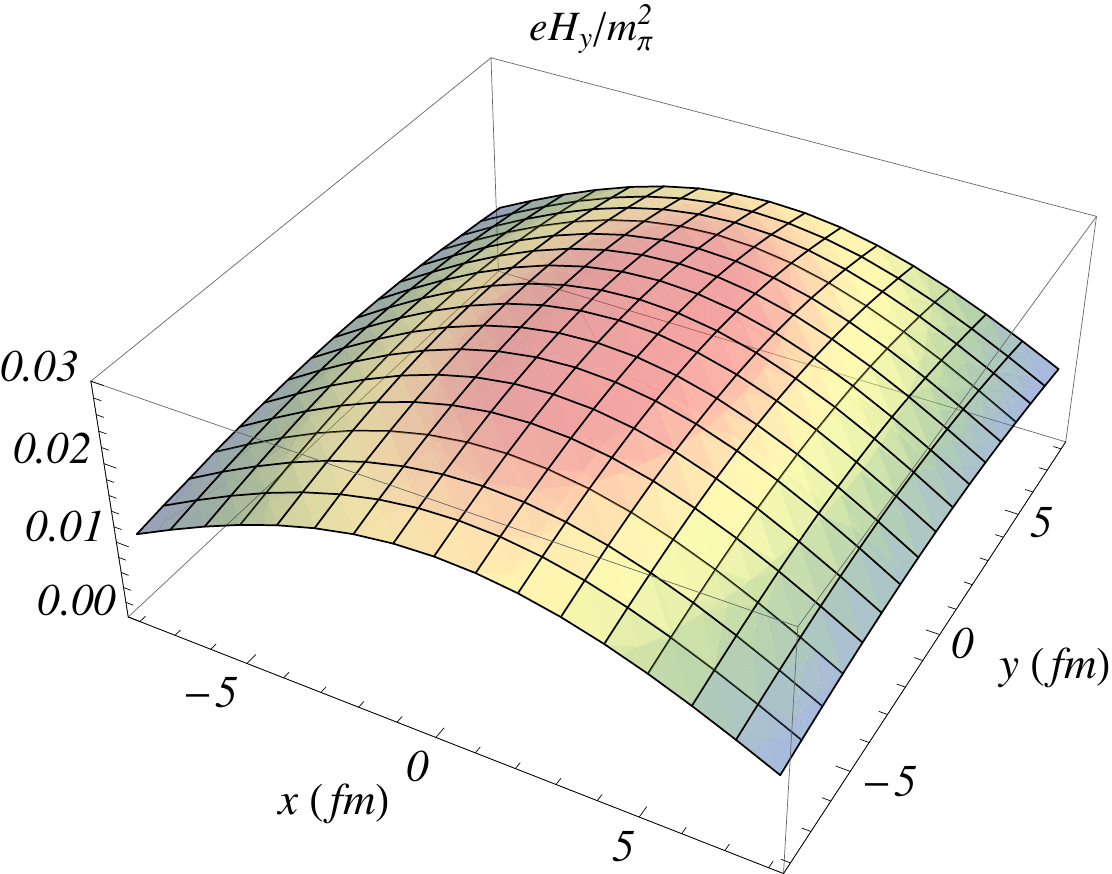} &
      \includegraphics[height=5.5cm]{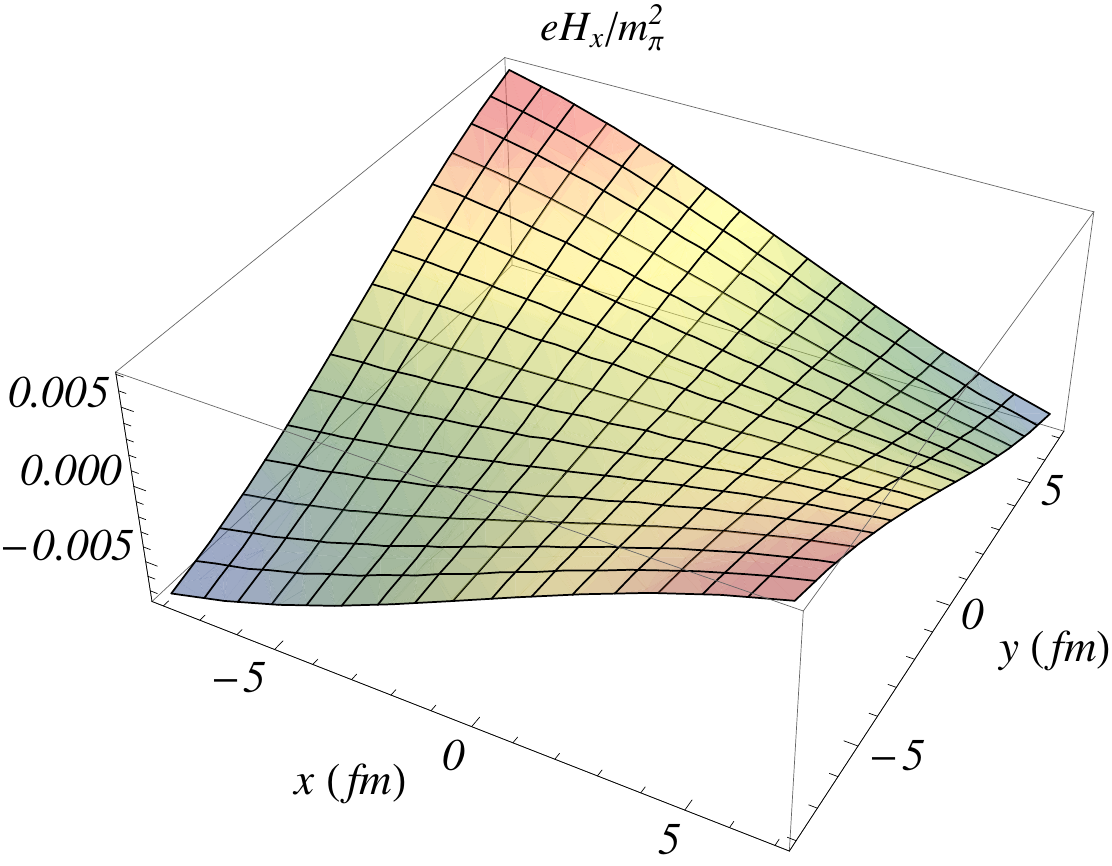}\\
      $(a)$ & $(b)$ \\
      \includegraphics[height=5.5cm]{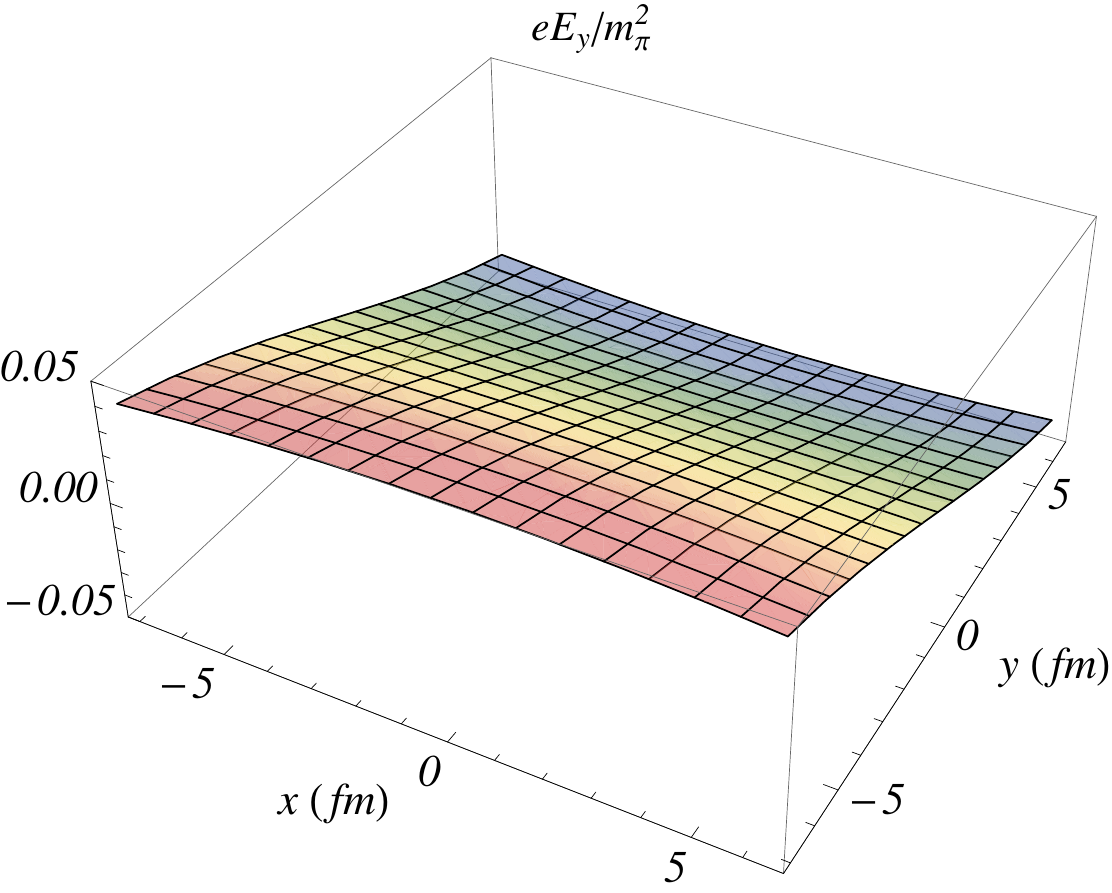} &
      \includegraphics[height=5.5cm]{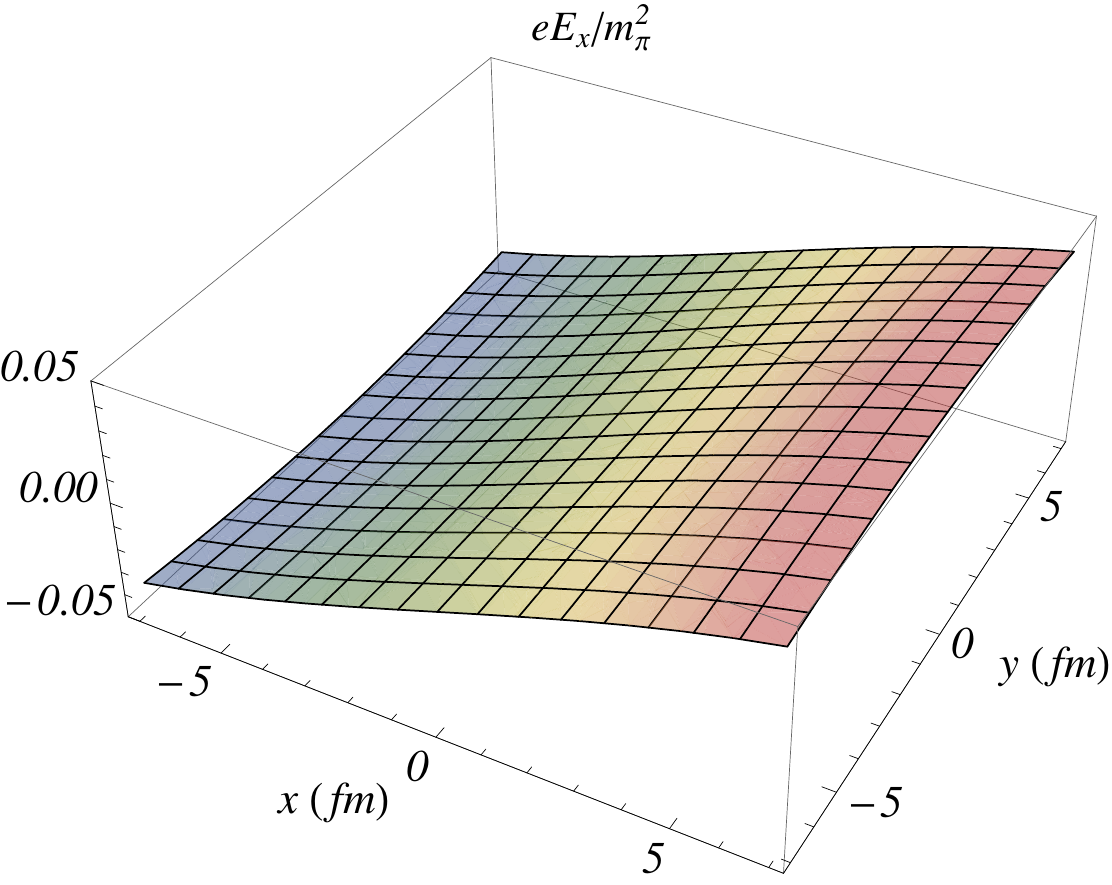}\\
      $(c)$ & $(d)$ 
      \end{tabular}
  \caption{ Structure of the field components in the transverse plane at mid-rapidity $z=0$, and $\gamma=100$, $B=7$~fm.  }
\label{fig:4}
\end{figure}
Space dependence is exhibited in \fig{fig:4}. We observe that the space variation of $H_y$ is mild. Other transverse components vary  more significantly as they are required to vanish at either $x=0$ or $y=0$ by symmetry. When averaged over the transverse plane, only $H_y$ component survives. However, one can think of observables sensitive to the field variations in the transverse plane.  

In summary, we presented exact analytical and numerical solution for space and time dependence of electromagnetic field produced in heavy-ion collisions. We confirmed our previous result \cite{Tuchin:2010vs} that nuclear matter plays a crucial role in its time-evolution.

\acknowledgments
This work  was supported in part by the U.S. Department of Energy under Grant No.\ DE-FG02-87ER40371.


\end{document}